\documentclass[referee]{raa}  

\usepackage{graphicx,times}             
\usepackage{natbib}
\usepackage{amssymb,amsmath}
\bibpunct{(}{)}{;}{a}{}{,}

\usepackage[driverfallback=dvipdfm,pagebackref=true]{hyperref}
\hypersetup{colorlinks = true, linkcolor = green, anchorcolor = red, citecolor = blue, filecolor = red, urlcolor = red}

\begin{document}
\title{Revisiting Pulsar Velocities Using Gaia Data Release 2}



   \volnopage{Vol.0 (20xx) No.0, 000--000}      
   \setcounter{page}{1}          

   \author{Meng Yang \inst{1,2},
           Shi Dai \inst{3,4},
           Di Li \inst{2,5,6},
           Chao-Wei Tsai \inst{2,5},
           Wei-Wei Zhu \inst{2,5}
           \and Jie Zhang \inst{7}
           }

   \institute{China West Normal University, Sichuan\ 637000, People's Republic of China;\\
        \and
            National Astronomical Observatories, Chinese Academy of Sciences, Beijing 100101, People's Republic of China;  {\it dili@nao.cas.cn}\\
        \and
            CSIRO Astronomy and Space Science, Australia Telescope National Facility, PO Box 76, Epping, NSW 1710, Australia; 
            {\it Shi.Dai@csiro.au}
            \\
        \and
            Western Sydney University, Locked Bag 1797, Penrith South DC, NSW 1797, Australia; \\
        \and
            University of Chinese Academy of Sciences, Beijing 100049, People's Republic of China;\\
        \and 
           NAOC-UKZN Computational Astrophysics Centre, University of KwaZulu-Natal, Durban 4000, South Africa;\\
        \and 
            Qilu Normal University, Shandong\ 250200, People's Republic of China;\\
\vs\no
   {\small Received 2021 February 2; accepted 2021 March 31}}

\abstract{
Precise measurements of neutron star (NS) velocities provide critical clues to the supernova physics and evolution of binary systems. Based on  Gaia Data Release 2 (DR2),  we selected a sample of 24 young ($<3$\,Myr) pulsars with precise parallax measurements and measured the velocity of their local standard of rest (LSR) and the velocity dispersion among their respective local stellar groups. The median velocity difference between thus calculated LSRs and the Galactic rotation model is $\sim 7.6$\,km\,s$^{-1}$, small compared to the typical velocity dispersion of $\sim 27.5$\,km\,s$^{-1}$. For pulsars off the Galactic plane, such differences grow significantly to  as large as $\sim40$\,km\,s$^{-1}$. More importantly, the velocity dispersion of stars in the local group of low-velocity pulsars can be comparable to their transverse velocities, suggesting that the intrinsic velocities of NS progenitors should be taken into account when we consider their natal kicks and binary evolution. We also examined the double NS systems J0737$-$3039A/B, and measured its transverse velocity to be $26^{+18}_{-13}$\,km\,s$^{-1}$ assuming nearby Gaia sources being representative of its birth environment. This work demonstrated the feasibility and importance of using Gaia data to study the velocity of individual systems and velocity distribution of NSs.
}

   \authorrunning{Y. Meng et al. }            
   \titlerunning{Revisiting Pulsar Velocities Using Gaia Data Release 2}  

   \maketitle
%
%
%
\section{Introduction}
\label{sect:intro}

Neutron stars (NSs) as a population are known to be moving faster than their progenitor stars since soon after discovery~\citep{Gunn1970}. A significant population of isolated NSs have velocities greater than 1000\,km\,s$^{-1}$~\citep[e.g.,][]{Cordes1993,Hobbs2005}. Such high velocities can be explained by natal kicks to NSs at their birth. Potential mechanisms to produce a natal kick include asymmetric explosion due to global hydrodynamical perturbations in the supernova core~\citep{Goldreich96, Burrows1996, Lai2000a, Lai2000b} and asymmetric neutrino emission in the presence of super-strong magnetic fields ($B>10^{15}$\,G) in the proto-neutron star~\citep[e.g.,][and references therein]{Lai1998,Arras1999a,Arras1999b}. \citet{bai89} pointed out that a large velocity can be produced by the disruption of massive binary systems triggered by an asymmetric kick of order 200\,km\,s$^{-1}$. A postnatal kick mechanism relying on asymmetric electromagnetic radiation from an off-centered dipole in a rapidly rotating pulsar has also been studied by \citet{Harrison1975}. As discussed by \citet{lcc01}, most kick mechanisms only predict a limited range of velocities, and therefore studies of velocity distribution of a large sample of NSs is important~\citep[e.g.,][]{acc02,Hobbs2005,Verbunt2017,Igoshev2020}.

While isolated NSs can have velocities up to $\sim1000$\,km\,s$^{-1}$, observations suggest that pulsars in binary systems and/or in star clusters have much smaller velocities. \citet{Tauris2017} show that in order to form some double NS (DNS) systems natal kicks have to be of order of tens km s$^{-1}$. \citet{tb96} carried out simulations for various millisecond pulsar evolutionary models and predicted the recoil velocity of such binary systems to be within a range of $\sim25$ to 200\,km\,s$^{-1}$. Velocities of pulsars in globular clusters have to be lower than the escape velocity of the clusters. Long-term timing observations of pulsars in 47 Tucanae show that measured pulsar velocities relative to the cluster are smaller than 24.5\,km\,s$^{-1}$~\citep{frk+17}, which is about half the escape velocity from the centre of the cluster~\citep{mam+06}. 

Studies of the velocity distribution of young isolated pulsars also show that a bimodal Maxwellian distribution with a low-velocity component is preferred~\citep[e.g.,][]{acc02,Igoshev2020}. For the low-velocity population and pulsars in binary systems, careful treatments of their respective local standard of rest (LSR) become important for a proper analysis of natal kick velocities of NSs. The Galactic rotation curve (GRC) has been used as a standard model for providing pulsar LSR \citep{Hobbs2005,Verbunt2017,Igoshev2020}. Obvious caveats of this approach include local deviation from the GRC, particularly away from the plane, and the velocity dispersion of stars (on the order of tens km\,s$^{-1}$), both of which can introduce systematic uncertainties. 

The data release 2 (DR2) of the Gaia satellite provides us the capability to study the dynamics of the Milky Way~\citep{Gaia2018a} and velocities of individual Galactic objects. It has already been used to study distances and velocities of binary pulsars~\citep{jkc+18}, and to search for companions to rotation-powered pulsars~\citep{ant21}. In this paper, we consider the feasibility and implication of using Gaia DR2 to study the LSR velocity and velocity dispersion of stars in the local group of known pulsars. We select a sample of young pulsars (including two binary systems PSRs B1259$-$63 and J2032$+$4127) and DNS PSR J0737$-$3039A/B, all of which have high precision parallax and proper motion measurements. Based on these information, we identify Gaia objects in their local group and then measure their LSR velocities and velocity dispersion. In Section~\ref{sec:data}, we describe the selection of pulsar sample and the method of measuring their LSR velocities. In Section~\ref{sec:result} we give our main results and discuss their implications. Conclusions will be given in Section~\ref{sec:conclusion}.

\section{Data and method}
\label{sec:data}
\subsection{The pulsar sample}
\label{subsec:psr}

We selected pulsars with precise measurements of parallax and proper motions from the Australian Telescope National Facility Pulsar Catalogue~\citep{Manchester2005}\footnote{version 1.64, \url{https://www.atnf.csiro.au/people/pulsar/psrcat}}. Only pulsars with better than three $\sigma$ measurements of their parallax ($\omega$) and proper motions ($\mu_{\alpha*}$ and $\mu_{\delta}$) were included. Most of these proper motions and parallaxes were measured using Very Long Baseline Interferometry (VLBI)~\citep[e.g.][]{chatterjee2009,Deller2019}, while the rest were based on optical observations of their companions~\citep[e.g.,][]{jkc+18}.

To ensure a proper reconstruction of the LSR representing the birth location of each pulsar, we further restricted our sample to young or low velocity pulsars. We selected pulsars with a characteristic age less than three million years (Myr) or transverse velocities less than 20\,km\,s$^{-1}$. These requirements enable us to select pulsars that are most likely still within 100\,pc from their birth locations. We note that a pulsar with a velocity of 20\,km\,s$^{-1}$ can travel a distance of $\sim20$\,pc in 1\,Myr.

\subsection{Gaia DR2}
\label{subsec:gaia}

The Gaia DR2 releases precise 3-dimensional position, parallax and proper motion measurements of $\sim1.3$ billion stars~\citep{Gaia2018b}. For sources brighter than $G=15$\,mag, uncertainties of parallax and proper motion from DR2 can reach 0.02\,mas and 0.07\,mas/yr, respectively. Same as for the pulsar sample, we only selected Gaia objects that meet the requirement of $\omega/\sigma_{\omega}>3$, where $\sigma_\omega$ is the uncertainty of Gaia parallax. The derived parallax distance can be less trustworthy when the uncertainty is large~\citep[e.g.,][]{ivc16}. Therefore, we only used distances with high reliability published by \citet{Jones2018} for Gaia DR2 sources.

\subsection{Measuring pulsar LSR velocities using Gaia DR2}
\label{subsec:lsr}

To provide a kinematic reference of each pulsar, selected Gaia objects within a distance of $R$ from a given pulsar are defined as stars in the same local group with the pulsar. We have tested a range of $R$ (from 10 to 50\,pc) in order to account for different sizes of the unknown local dynamical structures. Distant pulsars or pulsars far away from the Galactic plane have only a small number of Gaia objects in their local group, which are likely to be affected by high or low velocities of individual Gaia objects. Therefore, to avoid contaminants in the measurement, we only kept pulsars that have at least 20 nearby objects identified. As the result, a total of 22 isolated pulsars and three binary systems (PSRs B1259$-$63, J2032$+$4127 and J0737$-$3039A/B) are included in our analysis (see Table~\ref{tab:psr} for their details).

For each pulsar and Gaia objects within its local group, we iteratively rejected objects whose velocities are outside of three $\sigma$ until no further rejection is needed. We then calculated their median proper motion in the direction of right ascension RA ($\mu_{\alpha*,\rm{g}}$) and declination DEC ($\mu_{\delta,\rm{g}}$) in units of mas\,yr$^{-1}$. The transverse LSR velocity $V_{\rm Gaia}$, in units of km\,s$^{-1}$, is calculated as
\begin{equation}\label{vgaia}
V_{\rm Gaia} = 4.74\times D\times\sqrt{\mu_{\alpha*,\rm{g}}^2 + \mu_{\delta,\rm{g}}^2}\>\>,
\end{equation}
where $D$ is the pulsar distance from the Earth in units of kpc and 4.74 is a constant converting the unit from mas\,yr$^{-1}$ to km\,s$^{-1}$. Finally, we calculated pulsar transverse velocities relative to their local groups as
\begin{equation}
\label{vt}
V_{\perp} = 4.74\times D\times \sqrt{\mu_{\alpha*}^{2} + \mu_{\delta}^{2}}\>\>. 
\end{equation}
$\mu_{\alpha*}$ and $\mu_{\delta}$ are pulsar proper motions relative to their LSR defined as
\begin{equation}
\begin{split}
    \mu_{\alpha*} &= \mu_{\alpha*,\rm{psr}} - \mu_{\alpha*,\rm{g}}\\
    \mu_{\delta} &= \mu_{\delta,\rm{psr}} - \mu_{\delta,\rm{g}}\>\>,
\end{split}
\end{equation}
where $\mu_{\alpha*,\rm{psr}}$ and $\mu_{\delta,\rm{psr}}$ are the measured pulsar proper motions. To quantify the scatter in proper motion for a group of Gaia sources, we calculated the velocity dispersion of Gaia objects ($\sigma_{\rm Gaia}$, in units of km\,s$^{-1}$), which is defined as
\begin{equation}
\label{vd}
    \sigma_{\rm Gaia} = 4.74\times D\times \sqrt{\sigma_{\rm RA}^{2} + \sigma_{\rm DEC}^{2}} \>\>,
\end{equation}
where $\sigma_{\rm RA}$ and $\sigma_{\rm DEC}$ are the standard deviation of the proper motion in the RA and DEC directions, respectively.

\subsection{The Galactic rotation curve}

The velocity of pulsar LSR was estimated using the GRC in the literature~\citep[e.g.,][]{Verbunt2017,Igoshev2020}. For comparison , we calculated GRC-based LSR velocity, $V_{\rm GRC}$ (Table~\ref{tab:psr}). We used equations listed in the Appendix A of \citet{Verbunt2017} to convert the velocity of Galactic rotation and solar motion to the equatorial coordinate system. We adopted a flat Galaxy rotation curve with speed $v_{\rm R}=220$\,km\,s$^{-1}$ and distance of the Sun from the Galactic Centre $R_{\odot}=8.5$\,kpc. 

\begin{figure}[h]
  \begin{minipage}[t]{0.495\linewidth}
  \centering
   \includegraphics[width=70mm]{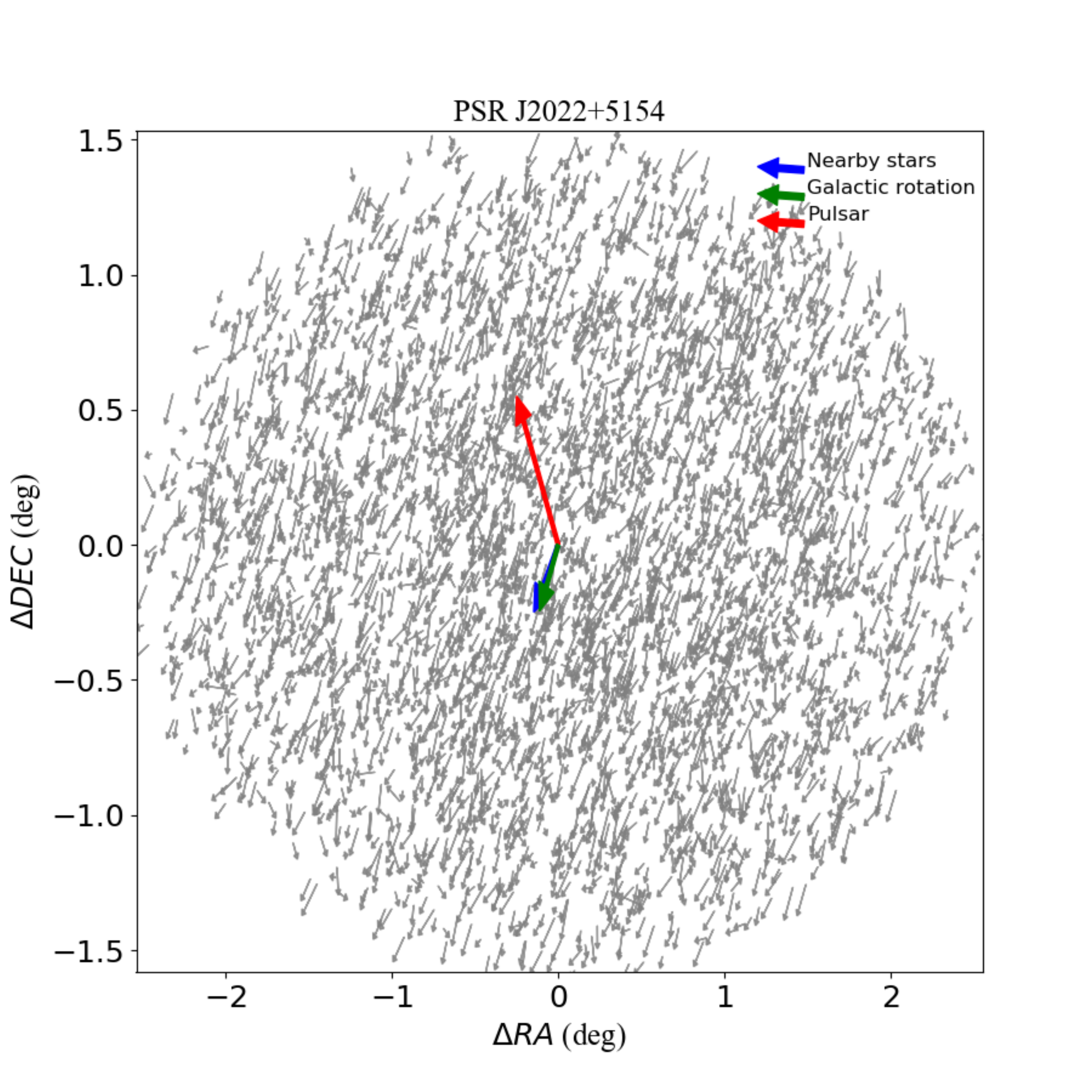}
   \end{minipage}
   \begin{minipage}[t]{0.495\textwidth}
  \centering
   \includegraphics[width=70mm]{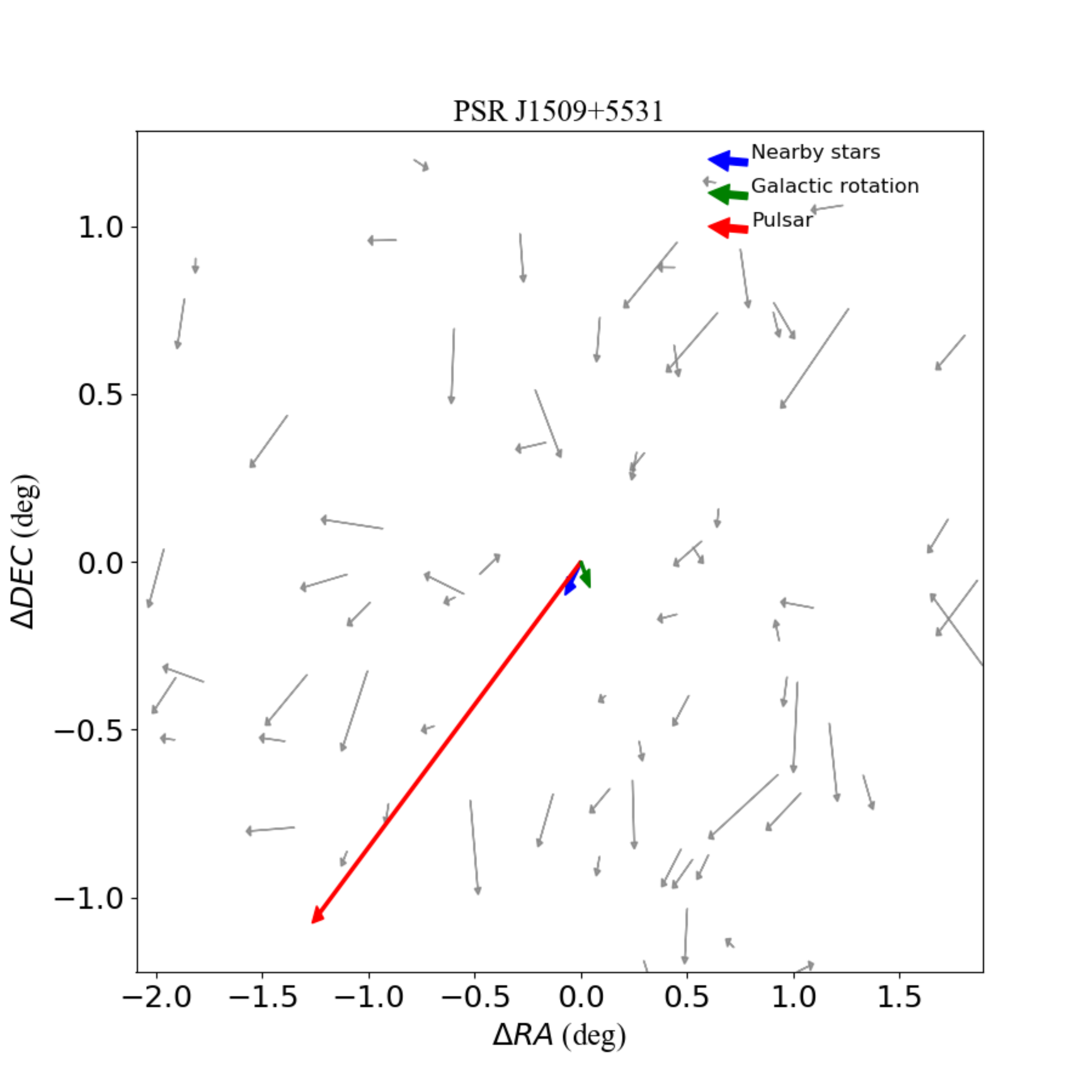}
  \end{minipage}
   \caption{\small Proper motion of Gaia objects in the local group of PSRs J2022$+$5154 (left) and J1509$+$5531 (right). Grey arrows show proper motions of Gaia objects. The red, blue and green arrows represent pulsar proper motion, mean proper motion of Gaia objects and prediction using the Galactic rotation curve, respectively. The length of red, blue and green arrows are scaled to be four times of those of grey arrows for PSR J2022$+$5154, and two times for PSR J1509$+$5531 }
  \label{fig:example}
\end{figure}

\section{Result and Discussion}
\label{sec:result}

\begin{figure}
\centering
\includegraphics[width=12cm]{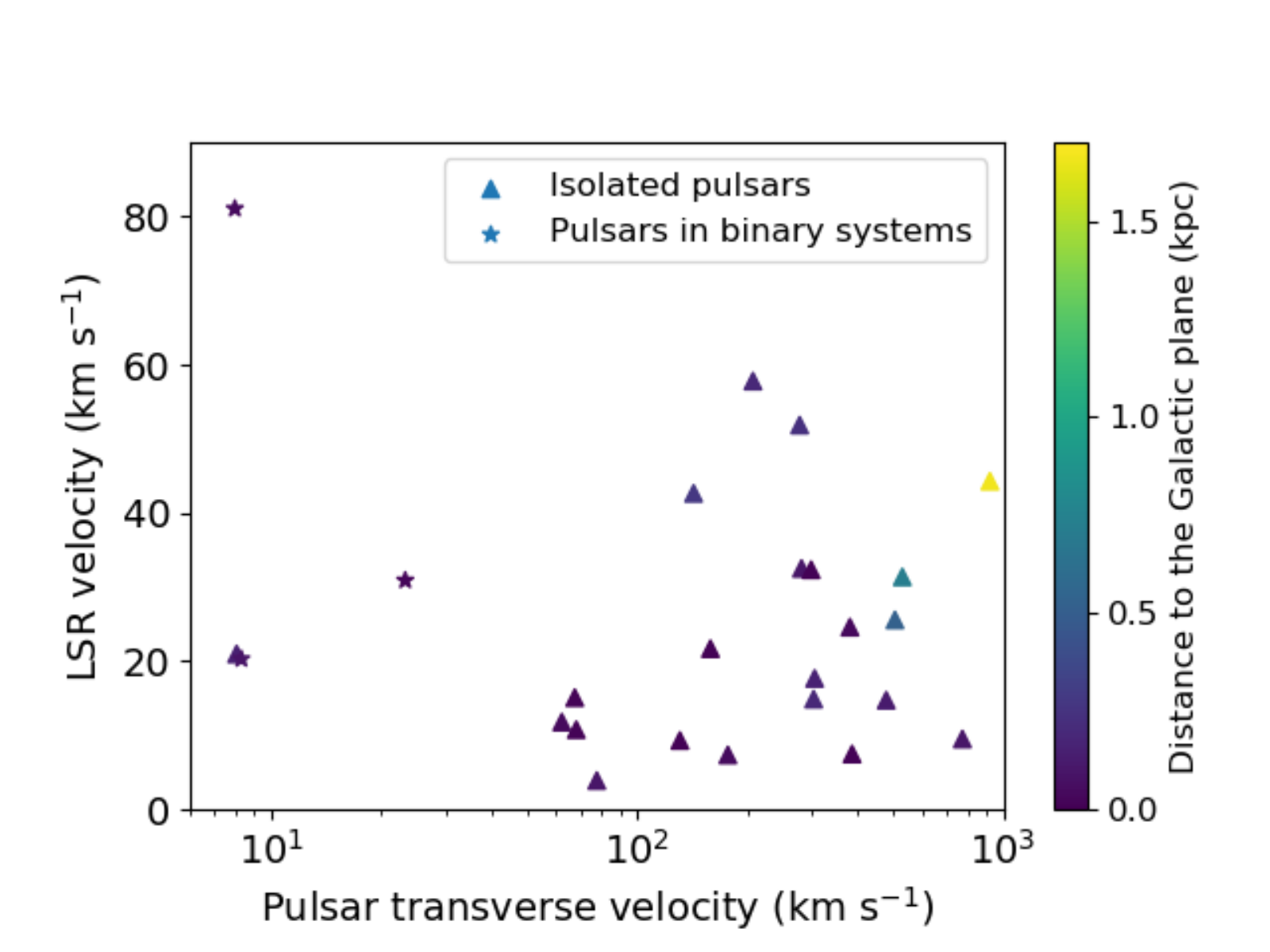}
\caption{Pulsar transverse velocity ($V_{\perp}$) versus velocity of pulsar LSR ($V_{\rm Gaia}$). $V_{\rm Gaia}$ and $V_{\perp}$ are calculated using Eq.~\ref{vgaia} and \ref{vt} as described in Section~\ref{subsec:lsr}. The colour bar shows the distance of the pulsar from the Galactic plane in units of kpc.}
\label{fig:relv_vgaia}
\end{figure}

We presented our measurements of transverse velocities and LSR velocities of 22 pulsars and the velocity dispersion of stars in their local group in Table~\ref{tab:psr}. In comparison, LSR velocities estimated using a GRC model are listed in the second last column of Table~\ref{tab:psr}. In Fig.~\ref{fig:example} we show two examples of Gaia sources identified in the local group of two known pulsars. Arrows and their length show the direction and scales of the proper motion of the surrounding stars within 50\,pc. For PSR J2022$+$5154, a clear coherent motion of the nearby stars is visible in the plot. This group movement agrees with the prediction based on the Galactic rotation curve. The proper motion of PSR J2022$+$5154 is significantly larger than the median proper motion of its local group (i.e., the LSR velocity) and is in a different direction with respect to the surrounding star group. The other pulsar, J1509$+$5531, is at the highest Galactic latitudes ($b=52.3^{\circ}$) in our sample. There are significant less Gaia sources identified in the local group of the pulsar due to the low stellar density away from the Galactic plane. The proper motions of these Gaia sources are much less coherent and show large scatters in both the RA and DEC directions. The LSR velocity of PSR J1509$+$5531 measured with Gaia DR2 is significantly different from the prediction of the Galactic rotation curve. The comparison between pulsar transverse velocities and their LSR velocities measured with Gaia DR2 are shown in Fig.~\ref{fig:relv_vgaia}. No clear correlation can be seen, consistent with the expectation of pulsars' velocity resulting from their natal kicks, which were presumably independent from either the GRC or the collective motion of the surrounding stellar groups.

\begin{figure}
\centering
\includegraphics[width=12cm]{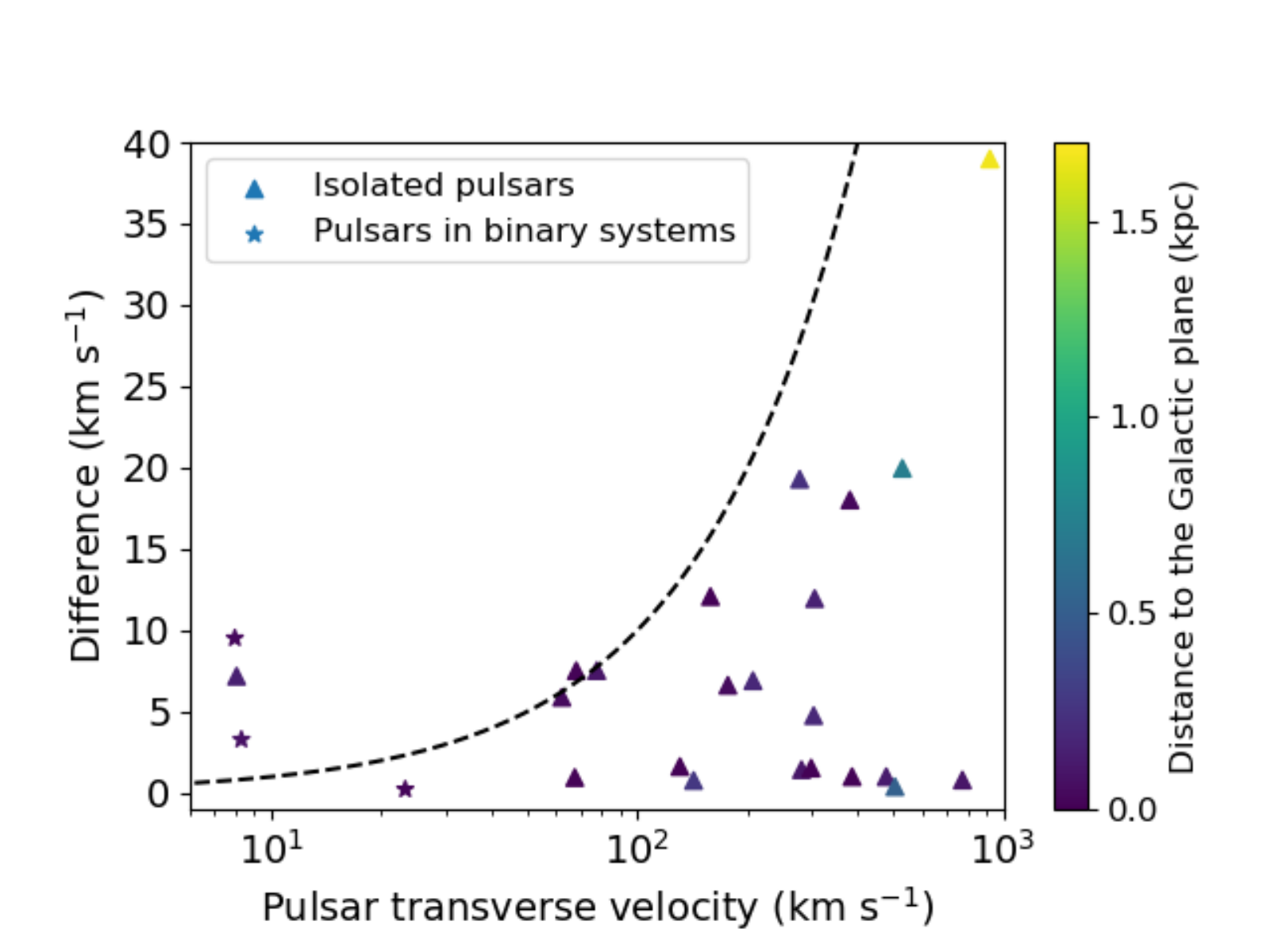}
\caption{The x-axis shows the pulsar transverse velocity ($V_{\perp}$) derived using Gaia DR2 (Eq.~\ref{vt}). The y-axis shows the absolute difference in pulsar transverse velocities derived using Gaia DR2 and a flat Galactic rotation curve. The dashed line shows the 10\% value of measured pulsar transverse velocities. The colour bar shows the distance of the pulsar from the Galactic plane in units of kpc.}
\label{fig:relv_diff}
\end{figure}

\begin{figure}
\centering
\includegraphics[width=12cm]{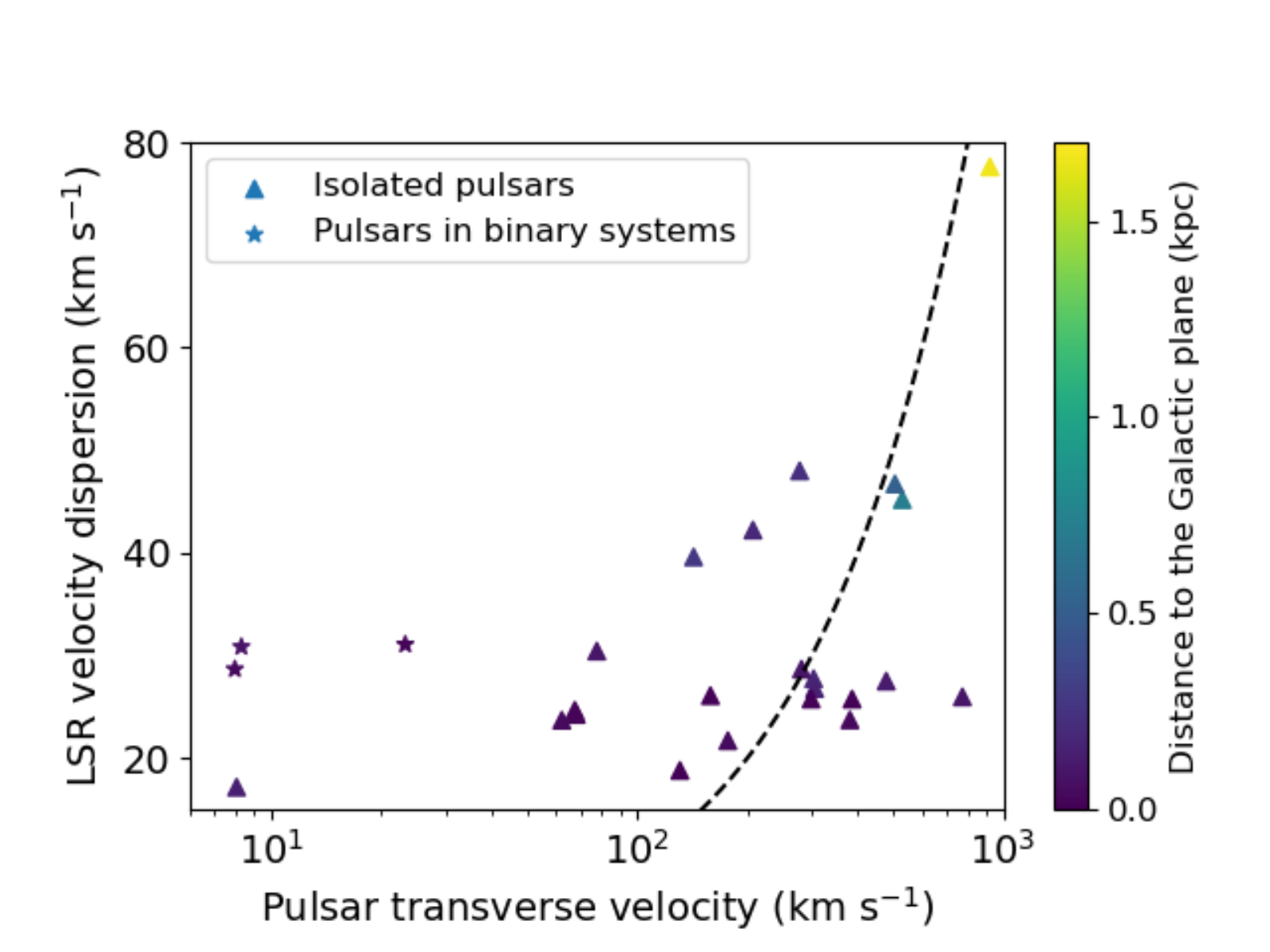}
\caption{The x-axis shows the pulsar transverse velocity ($V_{\perp}$) derived using Gaia DR2 (Eq.~\ref{vt}). The y-axis shows the velocity dispersion of pulsar LSR obtained with Gaia DR2 (Eq.~\ref{vd}). The dashed line shows the 10\% value of measured pulsar transverse velocities. The colour bar shows the distance of the pulsar from the Galactic plane in units of kpc.}
\label{fig:relv_disp}
\end{figure}

Previous studies accounted for the velocity of pulsar LSR simply using a flat Galactic rotation curve~\citep{Hansen1997,Hobbs2005,Verbunt2017,Igoshev2020}. Such an approach introduces systematic uncertainties into pulsar transverse velocities depending on their Galactic latitudes since a flat rotation curve model is only valid for the thin disc~\citep[e.g.,][]{rbf+17}. Another caveat of previous studies is the disregard of velocity dispersion among stars in the local stellar group surrounding the pulsar. In Fig.~\ref{fig:relv_diff} we show the absolute difference between pulsar transverse velocity derived using Gaia DR2 and a flat Galactic rotation curve. For the majority of pulsars in our sample, differences are of the order of $\sim10$\,km\,s$^{-1}$ and are generally much smaller than velocity dispersion and pulsar transverse velocities (the dashed line shows the 10\% value of measured pulsar transverse velocity). However, for pulsars with relatively low velocities (binary systems in particular), even though differences shown in Fig.~\ref{fig:relv_diff} are small compared with the velocity dispersion, our results show that the discrepancy is comparable to their transverse velocities and therefore using a flat GRC model could still affect our analysis of pulsar velocities. In Fig.~\ref{fig:relv_disp} we compare the velocity dispersion with pulsar transverse velocities. 15 out of 25 pulsars in our sample have a velocity dispersion larger than 10\% of their transverse velocities. For example, the LSR velocity dispersion of PSR J0614$+$2229 is measured to be 17.2\,km\,s$^{-1}$, much larger than its transverse velocity of $8\pm1$\,km\,s$^{-1}$. Since the characteristic age of this pulsar is only $\sim0.1$\,Myr, we expect it to be within 5\,pc from its birth location. The fact that the LSR velocity dispersion is larger than the pulsar velocity suggests that the natal kick velocity cannot be directly derived from the transverse velocity. The velocity of its progenitor needs to be taken into account.

It is well known that the velocity dispersion increases as a function of the distance from the Galactic plane~\citep[e.g.,][]{mcm12,bin12}. Although our sample of 25 pulsars is rather limited and most of them are close to the Galactic plane, we see that the velocity dispersion of pulsars away from the Galactic plane is significantly larger. For example, PSR J1509$+$5531, which is $\sim1.7$\,kpc away from the Galactic plane, has a velocity dispersion of $\sim80$\,km\,s$^{-1}$. As more pulsars will be discovered by future pulsar surveys~\citep[e.g.,][]{lwq+18,li19} and radio continuum surveys~\citep{djb+16,djh17} at relatively high Galactic latitudes, our results suggest that Gaia data of the ambient stars of the pulsar, rather than a simple Galactic rotation model, should be used for future studies of pulsar velocities. It is also worth noting that high-velocity pulsars with large distances from the Galactic plane could have travelled a substantial distance from their birth places, and therefore it becomes particularly important to treat their LSR velocities carefully and a  naive Galactic rotation model is insufficient.

Reliable measurements of velocities of DNS systems are crucial for understanding their evolutionary history~\citep[e.g.,][]{Tauris2017}. Previously, \citet{Tauris2017} used the Galactic model of \citet{McMillan2017} and Monte Carlo simulations to estimate the velocity of DNS LSR and its dispersion. Gaia data offer us the opportunity to measure DNS velocities relative to their LSR, although large uncertainties in their ages make it challenging to confidently identify their birth LSR. Here we consider a DNS system, PSR J0737$-$3039A/B, which has one of the lowest velocity in DNS systems and has precise measurements of its parallax and proper motions. Assuming that PSR J0737$-$3039A/B did not travel more than 50\,pc during its life time, thus the velocity dispersion of Gaia sources can be used to reconstruct the probability density function (PDF) of its progenitor system's velocity. We derived such a PDF of the transverse velocity (relative to the LSR), the median and $1\sigma$ uncertainty of which is $26^{+18}_{-13}$\,km\,s$^{-1}$, smaller but within uncertainty of that used in \citet{Tauris2017}. While to be treated with caution, such a result demonstrate the potential value of such analysis.

\begin{figure}[h]
    \begin{minipage}[t]{0.495\linewidth}
    \centering
    \includegraphics[width=70mm]{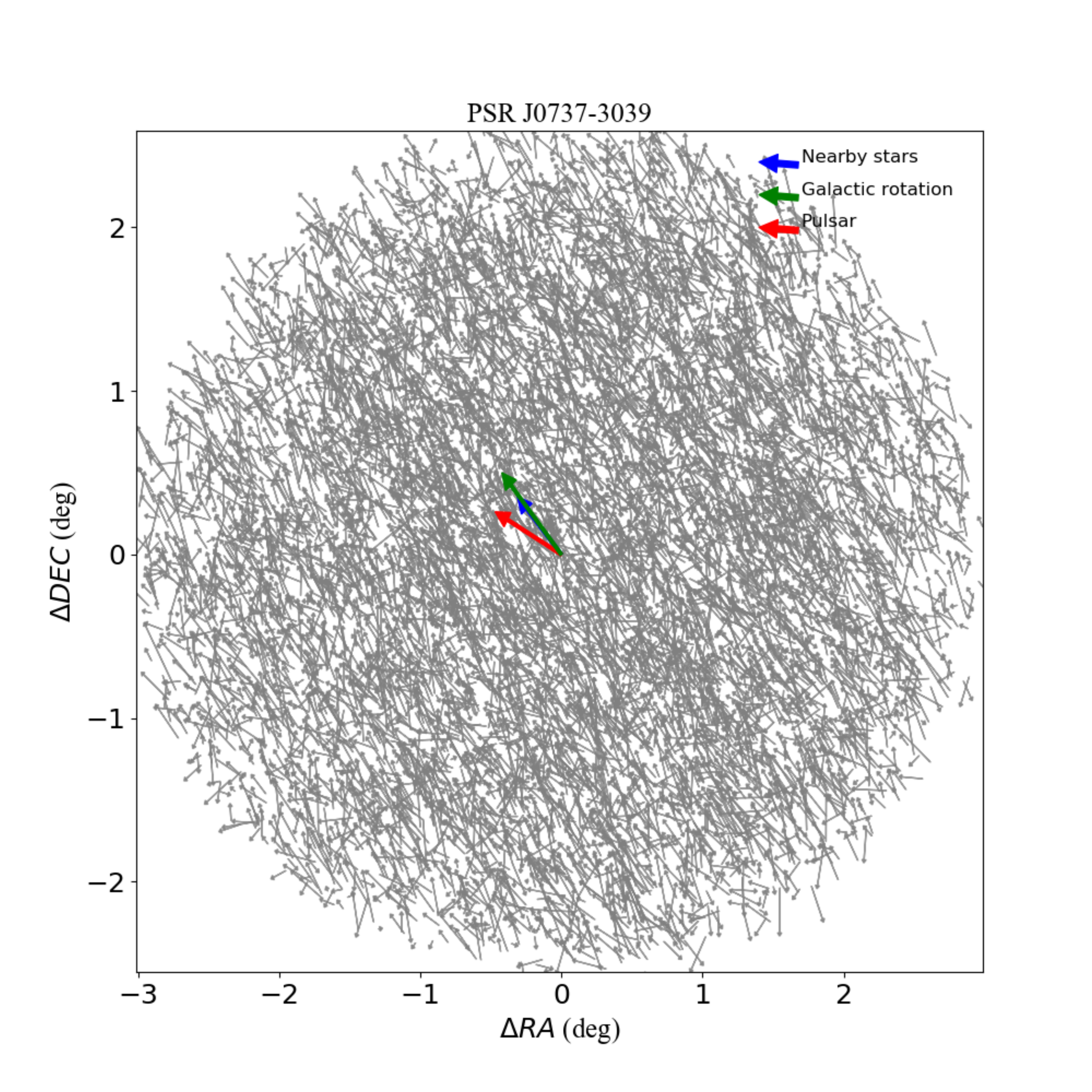}
    \end{minipage}%
    \begin{minipage}[t]{0.495\textwidth}
    \centering
    \includegraphics[width=70mm]{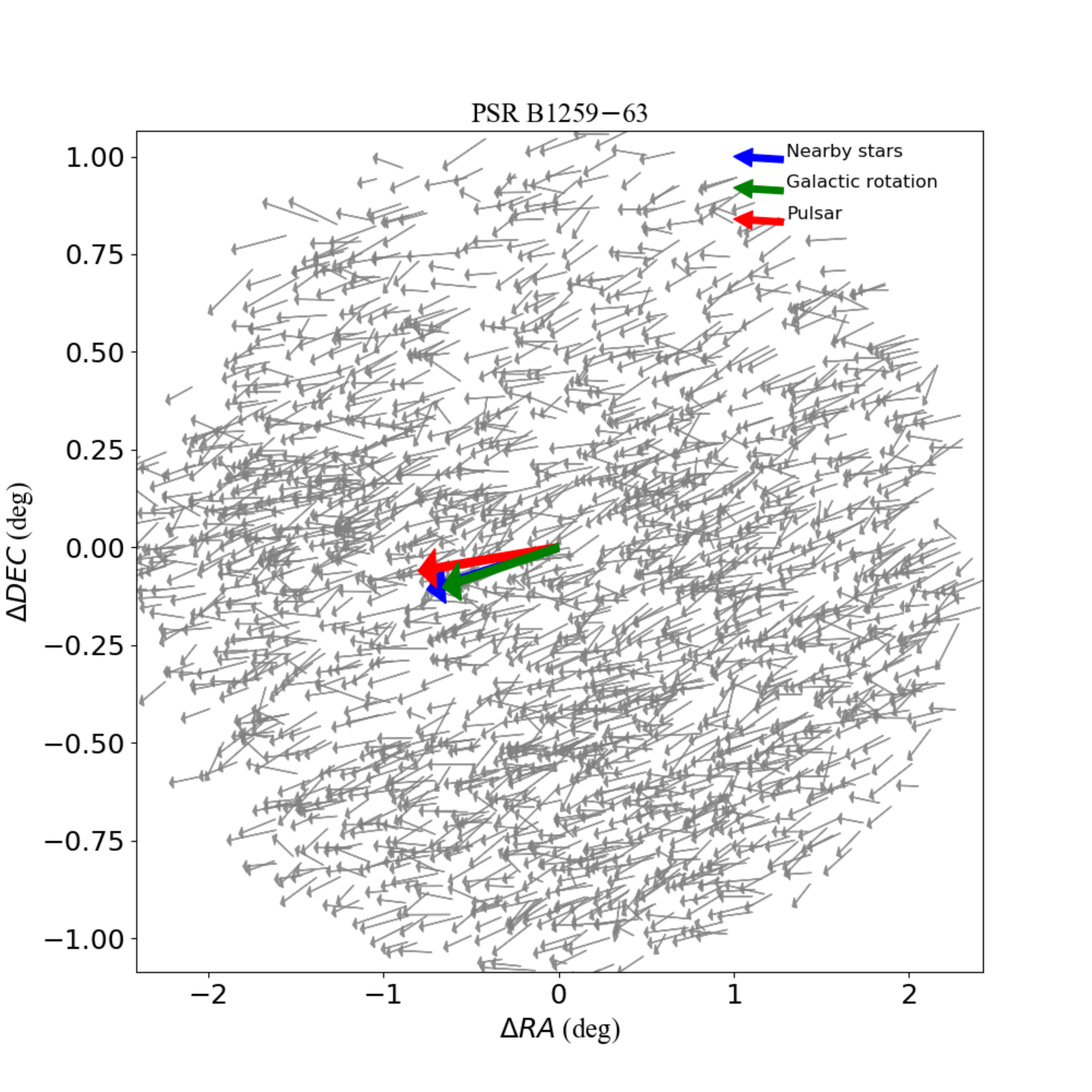}
    \end{minipage}
    \begin{minipage}[t]{0.495\textwidth}
    \centering
    \includegraphics[width=70mm]{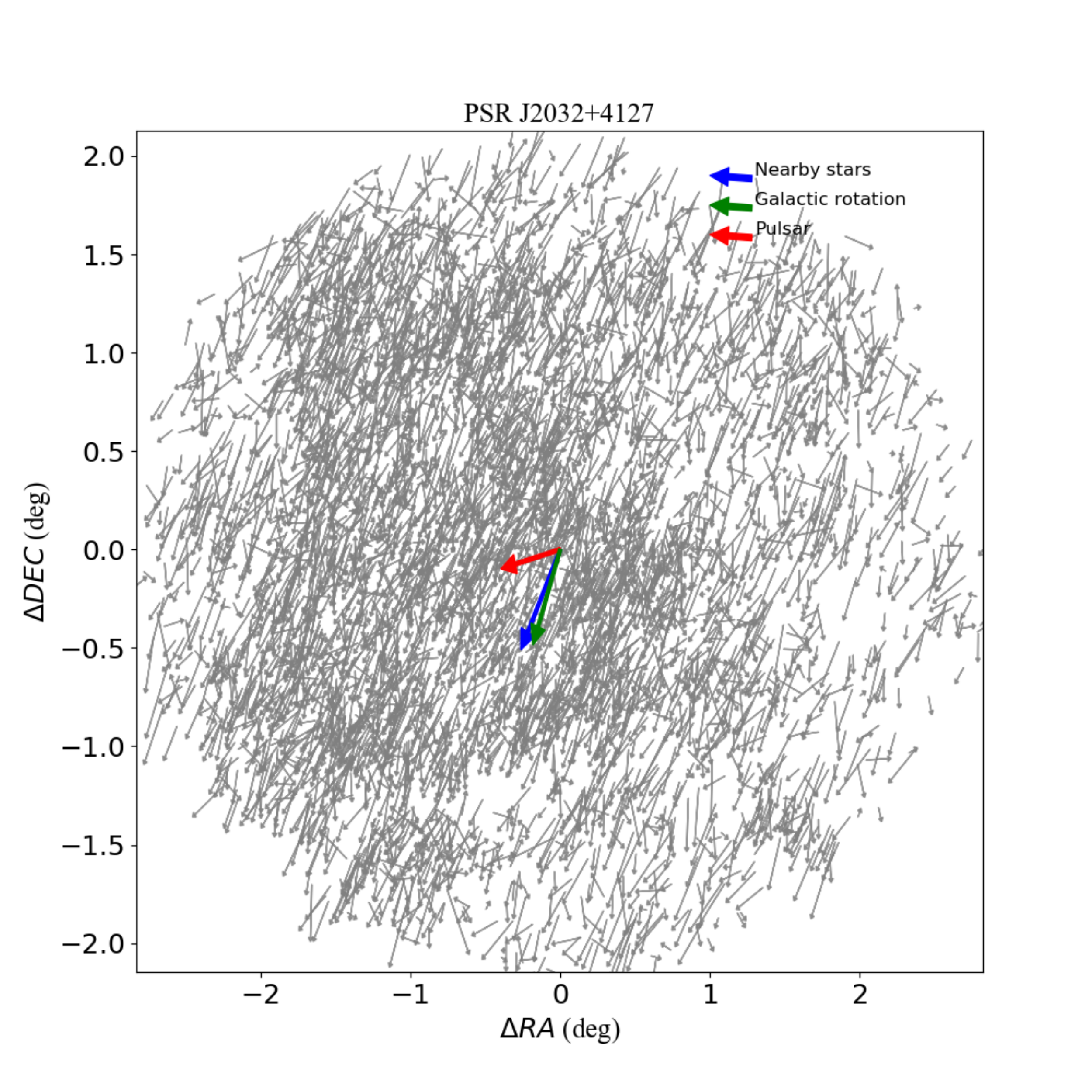}
    \end{minipage}
  \caption{Proper motion of Gaia objects in the local group of PSRs J0737$-$3039A/B, B1259$-$63 and J2032$+$4127. Grey arrows show proper motions of Gaia objects. The red, blue and green arrows represent pulsar proper motion, mean proper motion of Gaia objects and prediction using the Galactic rotation curve, respectively. The length of red, blue and green arrows are scaled to be four times of those of grey arrows.}
  \label{fig:binarypsr}
\end{figure}

Unlike PSR J0737$-$3039A/B, PSRs B1259$-$63 and J2032$+$4127 are young pulsars in binary systems with massive main sequence star companions~\citep{jml+92,lsk+15}, and are believed to be associated with the Cen OB1 association and the Cyg OB2 association, respectively. Since both B1259$-$63 and J2032$+$4127 are young, observed LSR velocity dispersion can be treated as an indicator of the velocity distribution of their progenitors (see Fig.~\ref{fig:binarypsr} for Gaia objects in their local group). Taking this into account, we obtained median transverse velocities (relative to their LSR) and their $1\sigma$ uncertainties for PSRs B1259$-$63 and J2032$+$4127 to be $23^{+17}_{-12}$\,km\,s$^{-1}$ and $32^{+20}_{-16}$\,km\,s$^{-1}$, respectively. These $1\sigma$ uncertainties are much larger than measurement uncertainties quoted in Table~\ref{tab:psr}, and should be considered for studies of the formation and evolutionary history of these two systems~\citep[see discussions in][]{Shannon2014,Miller-Jones2018}.
We measured the LSR velocity of PSR B1259$-$63 to be $\mu_{\alpha*,\rm{g}}=-6.6$\,mas\,yr$^{-1}$, $\mu_{\delta,\rm{g}}=-0.95$\,mas\,yr$^{-1}$. These values are slightly different from those published by \citet{co13} and agree with their measurements before removing field stars. This suggests that our measurements could be contaminated by field stars, but since these differences are well below the velocity dispersion of $\sim2.4$\,mas\,yr$^{-1}$,  transverse velocities presented above should not be much affected.

The local environment is different for different pulsars. It has been shown that the velocity dispersion of stars in the solar neighbourhood thin disc increases with time after star formation~\citep[][]{sg07}. This suggests that the observed velocity dispersion can be dominated by low-mass stars. The situation in star clusters and associations can be even more complicated. Gravitational encounters within a stellar system will drive the velocities of stars towards a dynamical thermal equilibrium, therefore stars of different mass have the same kinetic energy but different velocity. However, \citet{wbd+16} carried out a high-precision proper motion study of the Cygnus OB2 association and found no strong dependence of velocity distribution on stellar masses. In this work we did not consider the masses of the ambient stars when we discuss velocity dispersion. While it is beyond the scope of this paper to study the velocity dispersion as a function of stellar mass for each pulsar, this can be achieved by combining Gaia data with spectroscopic information from other surveys. More careful treatment of the velocity dispersion should be considered in  future studies.

\section{Conclusion}
\label{sec:conclusion}
We measured LSR velocities using Gaia DR2 for a sample of 24 young ($<3$\,Myr) pulsars and the only known double-pulsar system PSR J0737$-$3039A/B. We show that using a simple Galactic rotation curve model and ignoring velocity dispersion of stars in pulsar local groups is problematic, especially for pulsars away from the Galactic plane and/or with relatively low transverse velocities. Gaia data provide us crucial information on the velocity distribution of NS progenitors, thus shedding lights into their natal kick velocities and binary evolutionary history. Taking velocity dispersion of pulsar-hosting local stellar groups into account, we refined transverse velocity relative to their LSR of PSRs B1259-63 and J2032$+$4127 to be $23^{+17}_{-12}$\,km\,s$^{-1}$ and $32^{+20}_{-16}$\,km\,s$^{-1}$, respectively.
We also examined the double NS systems J0737$-$3039A/B, and measured its transverse velocity to be $26^{+18}_{-13}$\,km\,s$^{-1}$. This work demonstrated the feasibility and importance of using Gaia data for studies of NS velocities.

\begin{acknowledgements}
The authors thank Andrei Igoshev, Matthew Bailes and George Hobbs for discussions and comments. This work was supported by National Natural Science Foundation of China (grant numbers 11988101, 11725313, U2031121). This work has made use of data from the European Space Agency (ESA) mission
{\it Gaia} (\url{https://www.cosmos.esa.int/gaia}), processed by the {\it Gaia}
Data Processing and Analysis Consortium (DPAC,
\url{https://www.cosmos.esa.int/web/gaia/dpac/consortium}). Funding for the DPAC
has been provided by national institutions, in particular the institutions
participating in the {\it Gaia} Multilateral Agreement.
\end{acknowledgements}

\clearpage
\begin{table}
\centering
\caption{Basic information and velocities of 22 isolated young pulsars and three pulsars in binary systems.}
\label{tab:psr}
\begin{tabular}{lccccccccc}
\hline 
\hline
Name & l & b & $\mu_{\alpha*}$ & $\mu_{\delta}$ & $\ D$ & $V_{\perp}$ & $V_{\rm Gaia}$ & $V_{\rm GRC}$ & $\sigma_{\rm Gaia}$ \\
     & deg & deg & mas\,yr$^{-1}$ & mas\,yr$^{-1}$ & kpc & km\,s$^{-1}$ & km\,s$^{-1}$ & km\,s$^{-1}$ & km\,s$^{-1}$ \\
\hline
J0108+6608$^8$  & 124.6 & 3.3   & -32.75(3)  & 35.15(4)   & 2.14(15)  & 478.8(335)  & 14.7    & 19.8   & 27.5   \\
J0139+5814$^6$  & 129.2 & -4.0  & -19.11(7)  & -16.60(7)  & 2.65(30)  & 303.3(349)  & 14.9    & 21.9   & 27.7   \\
J0157+6212$^8$  & 130.6 & 0.3   & 1.57(6)    & 44.80(4)   & 1.8(1)    & 386.2(214)  & 7.5     & 13.0   & 25.7   \\
J0358+5413$^5$  & 148.2 & 0.8   & 9.3(5)     & 8.3(8)     & 1.04(18)  & 67.5(125)   & 15.1    & 8.3    & 24.6   \\
J0454+5543$^6$  & 152.6 & 7.5   & 53.34(6)   & -17.56(14) & 1.19(6)   & 305.2(166)  & 17.7    & 7.7    & 26.8   \\
                &       &       &            &            &           &             & &       &     \\
J0538+2817$^6$  & 179.7 & -1.7  & -23.57(10) & 52.87(10)  & 1.30(19)  & 381.0(556)  & 24.6    & 8.7    & 23.7   \\
J0614+2229$^8$  & 188.8 & 2.4   & -0.24(4)   & -1.25(4)   & 3.55(35)  & 8(1)     & 21.0    & 8.4    & 17.2   \\
J0630-2834$^7$  & 237.0 & -16.8 & -46.30(99) & 21.26(52)  & 0.332(46) & 77.5(108)   & 3.9     & 11.7   & 30.4   \\
J0633+1746$^3$  & 195.1 & 4.3   & 138(4)     & 97(4)      & 0.157(46) & 130.8(388)  & 9.3     & 6.9    & 18.8   \\
J0659+1414$^2$  & 201.1 & 8.3   & 44.07(63)  & -2.40(29)  & 0.288(30) & 62.2(65)    & 11.8    & 9.0    & 23.7   \\
                &       &       &            &            &           &             & &       &     \\
J0720-3125$^{12}$  & 244.2 & -8.2  & -93.9(12)  & 52.8(13)   & 0.36(13)  & 176.8(638)  & 7.3     & 14.1   & 21.7   \\
J0729-1836$^8$  & 233.8 & -0.3  & -13.06(11) & 13.27(44)  & 2.04(36)  & 158.5(285)  & 21.7    & 33.7   & 26.1   \\
J0835-4510$^9$  & 263.6 & -2.8  & -49.68(6)  & 29.9(1)    & 0.287(18) & 68.1(42)    & 10.7    & 18.3   & 24.2   \\
J0922+0638$^4$  & 225.4 & 36.4  & 18.8(9)    & 86.4(7)    & 1.21(19)  & 529.9(833)  & 31.4    & 25.6   & 45.2   \\
J1509+5531$^6$  & 91.3  & 52.3  & -73.64(5)  & -62.65(9)  & 2.10(13)  & 919.2(590)  & 44.3    & 22.7   & 77.6   \\
                &       &       &            &            &           &             & &       &     \\
J1820-0427$^8$  & 25.5  & 4.7   & -7.31(7)   & 15.89(8)   & 2.85(43)  & 277.7(424)  & 51.8    & 25.4   & 48.0   \\
J1833-0338$^8$  & 27.7  & 2.3   & -17.34(9)  & 15.0(3)    & 2.45(37)  & 281.0(430)  & 32.5    & 22.8   & 28.7   \\
J2022+2854$^1$  & 68.9  & -4.7  & -4.4(5)    & -23.6(3)   & 2.3(8)    & 207.0(720)  & 57.8    & 50.0   & 42.2   \\
J2022+5154$^1$  & 87.9  & 8.4   & -5.23(17)  & 11.5(3)    & 1.9(2)    & 142.5(189)  & 42.6    & 38.3   & 39.6   \\
J2048-1616$^6$  & 30.5  & -33.1 & 113.16(2)  & -4.60(28)  & 0.95(2)   & 505.8(106)  & 25.6    & 8.9    & 46.7   \\
                &       &       &            &            &           &             & &       &     \\
J2225+6535$^8$  & 108.6 & 6.8   & 147.23(23) & 126.5(1)   & 0.83(13)  & 773.1(1257) & 9.5     & 9.6    & 25.9   \\
J2354+6155$^8$  & 116.2 & -0.2  & 22.76(5)   & 4.897(25)  & 2.42(22)  & 298.6(277)  & 32.3    & 30.8   & 25.7   \\
\hline
\multicolumn{9}{c}{Young Pulsars in Binary Systems}\\
\hline  
B1259$-$63$^{13}$    & 304.2 & -1.0  & -7.01(3)   & -0.53(3)   & 2.6(3)    & 8.0(11)     & 81.0    & 70.2   & 28.7   \\
J2032+4127$^{11}$    & 80.2  & 1.0   & -2.991(48) & -0.742(55) & 1.38(6)   & 23.2(11)    & 30.9    & 27.4   & 31.1    \\
\hline
\multicolumn{9}{c}{Double NS Systems}\\
\hline   
J0737-3039A/B$^{10}$ & 245.2 & -4.5  & -3.82(62)  & 2.13(23)   & 1.15(19)  & 8.3(33)     & 20.3    & 30.2   & 30.8  \\
\hline
\end{tabular}

\begin{flushleft}
References for pulsar distances: (1) \cite{brisken2002}, (2) \cite{brisken2003}, (3) \cite{caraveo1996}, (4) \cite{chatterjee2001}, (5) \cite{chatterjee2004}, (6) \cite{chatterjee2009}, (7) \cite{deller_tbr2009}a, (8) \cite{Deller2019}, (9)\cite{dodson2003}, (10) \cite{deller_bt2009}b, (11) \cite{jkc+18}, (12) \cite{kaplan2007}, (13) \cite{Miller-Jones2018} \\
\end{flushleft}

\end{table}

\clearpage
\bibliographystyle{raa}
\bibliography{ref}

\label{lastpage}

\end{document}